\documentclass[prd,%showpacs,
preprintnumbers,nofootinbib,amsmath,eqsecnum]{revtex4}
  \usepackage{amssymb}
   \usepackage{amsmath}
    \usepackage{epsfig}
     \usepackage{bm}% bold math
      \usepackage{pifont}
\def\Li{\relax\ifmmode{\textbf{Li}_{2}}\else{Li$_2${ }}\fi}

\usepackage[usenames,dvipsnames]{xcolor}
\usepackage{refstyle}
\usepackage{fancybox}
\usepackage{amsmath}
\usepackage{empheq}
\usepackage{textpos}
\usepackage{pgfplots}
\usepackage{tikz}
\usepackage{dsfont}
\usepackage[nice]{nicefrac}
\usepackage{float}
\usepackage{enumerate}
\usetikzlibrary{calc}
\usetikzlibrary{shapes,arrows}
\usepackage{slashed}
\usetikzlibrary{calc,matrix,arrows,decorations.pathmorphing,decorations.markings,trees,fadings,decorations.pathreplacing}
\usetikzlibrary{trees}
\usetikzlibrary{intersections,shapes.arrows}
\usepackage{wrapfig}

\makeatletter

\AtBeginDocument{}
\AtBeginDocument{}
\AtBeginDocument{}
\newcommand{\lyxmathsym}[1]{\ifmmode\begingroup\def\b@ld{bold}
  \text{\ifx\math@version\b@ld\bfseries\fi#1}\endgroup\else#1\fi}

\RS@ifundefined{subref}
  {\def\RSsubtxt{section~}\newref{sub}{name = \RSsubtxt}}
  {}
\RS@ifundefined{thmref}
  {\def\RSthmtxt{theorem~}\newref{thm}{name = \RSthmtxt}}
  {}
\RS@ifundefined{lemref}
  {\def\RSlemtxt{lemma~}\newref{lem}{name = \RSlemtxt}}
  {}

\newcommand{\half}{\frac{1}{2}}

\newcommand{\es}{\frac{1}{\e^2}}

\newcommand{\be}{\begin{equation}}
\newcommand{\ee}{\end{equation}}

\newcommand{\ba}{\begin{eqnarray}}
\newcommand{\ea}{\end{eqnarray}}
\newcommand{\bg}{\begin{gather}}
\newcommand{\foma}{\end{gather}}

\newcommand{\vecc}[1]{\mbox{\boldmath $#1$}}

\newcommand{\nn}{\nonumber}
\newcommand\shapeder{\Bigg\langle\frac{\delta}{\delta \ln S}\Bigg\rangle}
\newcommand\shapedertilde{\Bigg\langle\frac{\delta}{\delta \ln \tilde S}\Bigg\rangle}

\newcommand\lr[1]{{\left({#1}\right)}}
\newcommand\lrb[1]{{\left\lbrace{#1}\right\rbrace}}

\newcommand{\gcusp}{\G_\text{cusp}}
\newcommand \widebar [1] {\overline{#1}}

\def \vep {\epsilon}

\newcommand\alpi{\frac{\a_s} {\pi}}

\newtheorem{theorem}{Theorem}[section]
\newtheorem{lemma}[theorem]{Lemma}
\newtheorem{proposition}[theorem]{Proposition}
\newtheorem{corollary}[theorem]{Corollary}
\newtheorem{definition}[theorem]{Definition}
\newtheorem{example}[theorem]{Example}

\newcommand{\qed}{\nobreak \ifvmode \relax \else
      \ifdim\lastskip<1.5em \hskip-\lastskip
      \hskip1.5em plus0em minus0.5em \fi \nobreak
      \vrule height0.75em width0.5em depth0.25em\fi}

\newcommand{\bd}{\begin{definition}}
\newcommand{\ed}{\end{definition}}
\newcommand{\bl}{\begin{lemma}}
\newcommand{\el}{\end{lemma}}
\newcommand{\bt}{\begin{theorem}}
\newcommand{\et}{\end{theorem}}
\newcommand{\bc}{\begin{corollary}}
\newcommand{\ec}{\end{corollary}}
\newcommand{\bex}{\begin{example}}
\newcommand{\eex}{\end{example}}
\newcommand{\bp}{\begin{proposition}}
\newcommand{\ep}{\end{proposition}}
\newcommand{\baa}{\begin{align}}
\newcommand{\eaa}{\end{align}}

\def\pd{\partial}

\def\e{\epsilon}

\def\pd{\partial}

\def\<{\langle}
\def\>{\rangle}

\def\a{\alpha}

\def\g{\gamma}  \def\G{\Gamma}

\def\m{\mu}
\def\n{\nu}

\def\({\left(}
\def\[{\left[}
\def\){\right)}
\def\]{\right]}

\def\pd{\partial}

\tikzset{
		wilsonline/.style={draw, double distance=1.5pt, postaction={decorate}, decoration={markings,mark=at position .5 with 				 {\arrow{>}}}},
		wilsonline2/.style={draw=NavyBlue, double distance=2pt, postaction={decorate}}
		}
\tikzset{
		photon/.style={decorate, decoration={snake}, draw=red},
		electron/.style={draw=blue, postaction={decorate},decoration={markings,mark=at position .55 with {\arrow[draw=blue]{>}}}},
		gluon/.style={decorate, draw=magenta,     decoration={coil,amplitude=4pt, segment length=5pt}},
		proton/.style={draw, triple distance=2pt, post action{decorate},decoration={markings,mark=at position .55 with {\arrow[draw=blue]{>}}}},
							triplearrow/.style={
							  draw=black!75,
							  color=black!75,
							  thick,
							  double distance=4pt,
							  postaction={decorate},decoration={markings,mark=at position .55 with {\arrow[draw=blue]{>}}},
							  >=stealth},
							  thirdline/.style={draw=black!75, color=black!75, thick, postaction={decorate},decoration={markings,mark=at position .55 with {\arrow[draw=blue]{>}}}, >=stealth},
							triplearroww/.style={
							  draw=black!75,
							  color=black!75,
							  thick,
							  double distance=4pt,
							  postaction={decorate},decoration={markings,mark=at position .55 with {\arrow[draw=red]{>}}}},
							  thirdlinee/.style={draw=black!75, color=black!75, thick, preaction={decorate},decoration={markings,mark=at position .55 with {\arrow[draw=red]{>}}}}
		}
\definecolor{kyelloworange}   {RGB}{255, 210,  110}
\tikzset{
point/.style={minimum size=0pt, inner sep=0pt},
dot/.style={minimum size=1.5pt, inner sep=1.3pt,circle,draw=NavyBlue!95,fill=NavyBlue!95},
photon/.style={decorate, decoration={snake}, draw=red,very thick},
gluon/.style={decorate, draw=Maroon, very thick,  decoration={coil,amplitude=2pt, segment length=3pt}},
quark/.style={draw=blue,very thick, postaction={decorate},
	decoration={markings,mark=at position .55 with {\arrow[draw=blue]{>}}}},
antiquark/.style={draw=blue,very thick, postaction={decorate},
	decoration={markings,mark=at position .55 with {\arrow[draw=blue]{<}}}},
eikonal/.style={double,double distance=1.5pt,draw=NavyBlue!60!blue, postaction={decorate},
	decoration={markings,mark=at position .55 with {\arrow[draw=NavyBlue!70!blue]{>}}}},
antieikonal/.style={double,double distance=1.5pt,draw=NavyBlue!60!blue, postaction={decorate},
	decoration={markings,mark=at position .55 with {\arrow[draw=NavyBlue!70!blue]{<}}}},
higgs/.style={draw=black,very thick, postaction={decorate},
	 decoration={markings,mark=at position .55 with {\arrow[draw=red]{>}}}}
}

\pgfkeys{
 	/marker/.is family, /marker,
	default/.style = {color=LimeGreen,inner sep=1.5mm},
	default2/.style = {color=LimeGreen,inner sep=0.5pt},
	color/.estore in = \markercolor,
	inner sep/.estore in = \markerinnersep
}
}] (0,0) -- (1,2) -- (3,2) -- (2,0) -- cycle;
}
}] (0,0) -- (0,2) -- (2,2) -- (2,0) -- cycle;
}
}] (0,0) -- (2,0) -- (3,2) -- (1,2) -- cycle;
}

}] (0,0) -- (2,0) -- (2,2) -- (0,2) -- cycle;
}

\usepackage{pgfplots}
\usepackage{tikz}
\usetikzlibrary{trees}
\usetikzlibrary{decorations.pathmorphing}
\usetikzlibrary{decorations.markings}
\usetikzlibrary{shapes.arrows}
\usepackage{refstyle}
\usepackage{amsmath,tkz-euclide,tkz-fct}
\usetkzobj{all}

\begin{document}
\thispagestyle{empty}
\date{\today}

\title{On Geometric Scaling of Light-Like Wilson Polygons: Higher Orders in $\alpha_s$}

\author{I.O.~Cherednikov\footnote{On leave from: Bogoliubov Laboratory of Theoretical Physics, JINR, 141980 Dubna, Russia}}
\email{igor.cherednikov@uantwerpen.be}
\affiliation{Departement Fysica, Universiteit Antwerpen, B-2020 Antwerpen, Belgium\\}
%\affiliation{Bogoliubov Laboratory of Theoretical Physics, JINR, 141980 Dubna, Russia\\}
\author{T.~Mertens}
\email{tom.mertens@uantwerpen.be}
\affiliation{Departement Fysica, Universiteit Antwerpen, B-2020 Antwerpen, Belgium\\}
%\vspace {10mm}
%\cleardoublepage
\begin{abstract}
We address the scaling behaviour of contour-shape-dependent ultra-violet singularities of the light-like cusped Wilson loops in Yang-Mills and ${\cal N} = 4$ super-Yang-Mills theories in the higher orders of the perturbative expansion. We give the simple arguments to support the idea that identifying of a special type of non-local infinitesimal shape variations of the light-like Wilson polygons with the Fr\'echet differentials results in the combined geometric and renormalization-group evolution equation, which is applicable beyond the leading order exponentiated Wilson loops.
\end{abstract}
%\pacs{13.60.Hb,13.85.Hd,13.87.Fh,13.88.+e}
%Keywords: Wilson loops
%          Gauge invariance
%          parton distribution functions
%          Renormalization and anomalous dimensions
%          Phase factors
%          Single spin asymmetries
\maketitle

%\tableofcontents
%\newpage

\section{Introduction}
Wilson loops with light-like parts and obstructions (like cusps and/or self-intersections) occurs in a number of important gauge-invariant hadronic and vacuum correlation functions, among which transverse momentum and distance dependent parton distribution functions in the theory of strong interaction, multi-gluon scattering amplitudes in ${\cal N}= 4$ super-Yang-Mills theory, jet quenching and transverse-momentum broadening functions in QCD and AdS/CFT are worth mentioning (for details, see, e.g., Refs. \cite{KR87,LargeX_KM,Makeenko_LC_WL,CS_all,LC_TMD_all,BMM_all,CMTVDV_2013,Jet_all} and Refs. therein). Calculation of these correlation functions within {a} QFT setting calls for careful treatment of the emerging singularities (ultraviolet, infrared and rapidity). The structure of these divergences is normally more involved than the one of the fully non-light-like Wilson functionals that results in the highly non-trivial renormalization properties of the former \cite{WL_LC_all}.  On the other hand, the (partially) light-like Wilson loops can be treated as elements of a generalized loop space \cite{Tavares:1993pw}, for which the equations of motion govern their behaviour under shape variations. One has to take into account, however, that in the quantum field-theoretic setting the infinitesimal shape variations do not necessary imply infinitesimal variations of the Wilson exponentials defined on these paths. The issue of the emerging singularities arise here as well \cite{WL_Renorm}. Namely, in the class of smooth paths $\gamma$ the variations of the corresponding Wilson loops resulting from variations in the contours can be consistently described by the Makeenko-Migdal loop equations \cite{MM_WL,WL_Renorm,St_all}. In contrast, the analysis of the cusped contours possessing some light-like segments requires more careful approach to introduction of the shape variations because of the extra divergences, which affect the renormalization properties of the Wilson loops under consideration. 

\begin{figure}[h!]
			\includegraphics[width=.9\textwidth]{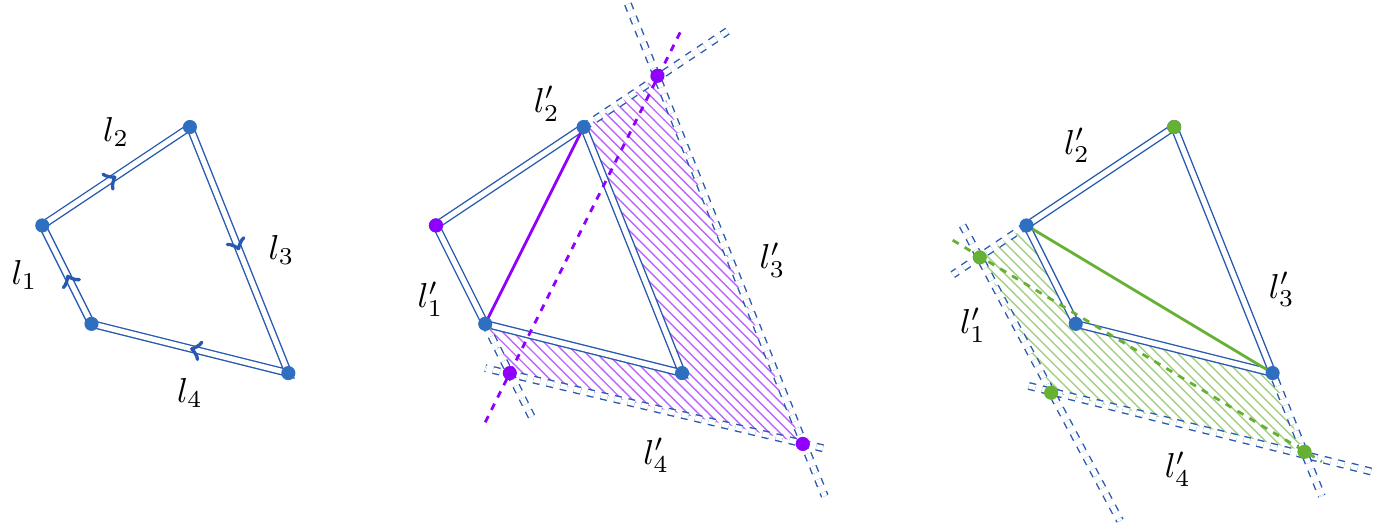}
			\caption{Quadrilateral  contour $\gamma$ with the light-like sides $\ell_i^2 =0$ (left panel); Examples of the shape variations generated by the differential operators (\ref{eq:newdiffop}) (right panels).}
			\label{fig:WLQ_possible_area_var_LC}
		\end{figure}

\begin{figure}[h!]
			\center
			\begin{tikzpicture}
				\node[point] at (0,0) {};
					\begin{scope}[shift={(0,0)}, scale=.7]
						% defining the boundary points
						\tkzDefPoint(0,0){O}
						\tkzDefPoint(3,0){A}
						\tkzDefPoint(-2,-4){B}
						\tkzDefPoint(1,-4){C}
						% drawing the contour
						\draw[quark] (B) --node[left=.5cm] {$\ell_1$} (O);
						\draw[eikonal] (O) -- node[above=.5cm] {$\ell_2$} (A);
						\draw[quark] (A) -- node[right=.5cm] {$\ell_3$}(C);
						%\tkzDrawLines[add = 0 and .5](O,A O,B)
						\tkzDrawPoints(O,A)
						%\tkzLabelPoints[below](O,A,B)
					\end{scope}
			\end{tikzpicture}
			\caption{$\Pi$-shaped contour with light-like $\ell_2^2 = 0$ and non-light-like $\ell_{1,3}^2 \neq 0$ parts.}
			\label{fig:WLQ_pi_shape}
		\end{figure}
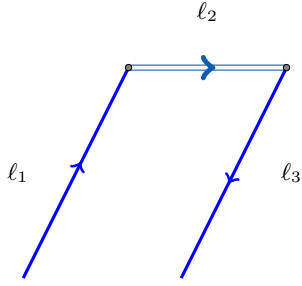

Recently \cite{WL_UA} we analysed the geometric and renormalization behaviour of the simplest object possessing the properties under consideration, i.e. the planar quadrilateral contour parametrized by the light-like vectors $\ell_i,\ i \in \lrb{1,2,3,4}$,  Fig. \ref{fig:WLQ_possible_area_var_LC}.
Note that this contour should by considered as a ``dual'' one to the ``original'' Wilson polygon on the light-cone. The latter, given that the lengths $\ell_i$ are allowed to be different, is obviously not planar. In what follows we are concerned with the properties of this planar dual version unless stated otherwise. 
We proposed a special class of the infinitesimal contour variations which are generated by the differential operators
		\begin{equation}\label{eq:newdiffop}
			S_{ij}\frac{\delta}{\delta S_{ij}}  = 
			 (2{\ell}_i \cdot \ell_j) \frac{\partial}{\partial  (2\ell_i \cdot \ell_j)} \ , \  
			S_{ij}   =  (\ell_i + \ell_j)^2 
			 \ ,		
			\end{equation}
			and
			\begin{equation}
			\Bigg\langle\frac{\delta}{\delta \ln S}\Bigg\rangle_1 
			=
			S_{12}\frac{\delta}{\delta S_{12}} +  S_{23}\frac{\delta}{\delta S_{23}} \ , \  \Bigg\langle\frac{\delta}{\delta \ln S}\Bigg\rangle_2 
			=
			S_{23}\frac{\delta}{\delta S_{23}} +  S_{34}\frac{\delta}{\delta S_{34}} \ , \ \text{etc.}  
			\label{eq:newdiffop_1}
			\end{equation}
where $S_{ij}$ are the variables which variations determine the shape changes preserving the ``angles'' of the dual quadrilateral planar contour, Fig. \ref{fig:WLQ_possible_area_var_LC}, with the sides 
\begin{equation}
  \ell_1^\mu = \ell_1 (1^+,0^-, \vecc 0_\perp) \ , \   \ell_2^\mu = \ell_2(0^+,1^-, \vecc 0_\perp)\ , \   \ell_3^\mu = - \ell_ (1^+,0^-, \vecc 0_\perp) \ , \    \ell_4^\mu = - \ell_2 (0^+,1^-, \vecc 0_\perp) \  ,       
\end{equation}
first studied in Refs. \cite{WL_LC_all}.  
The differential operator (\ref{eq:newdiffop_1}) does not, however, secure the shape scaling by default. The reason is that the singularities of the light-like Wilson polygons depend only on the Mandelstam variables $S_{ij}$, which do not determine the shape completely. In Ref. \cite{ChM_2014_1} we found that these operators arise as a particular case of 
Fr\'echet differential operator associated with a specific diffeomorphism-generating vector field.  We demonstrated explicitly that at the leading order in $\alpha_s$ the
operator (\ref{eq:newdiffop}) for $i=1, j=2$ coincides with the Fr\'echet derivative associated to a vector field defined as
$$
V_1^\m = \ell_1^\m+\ell_2^\m = (\ell_1^{+} \ , \ \ell_2^{-} \ , \ \vecc 0_\perp)
$$ 
generating the angle-conserving shape variations, Fig. \ref{fig:WLQ_possible_area_var_LC}  (see also Ref. \cite{Tavares:1993pw} for more details on this association). The logarithmic Fr\'echet derivative then reads
\be\label{eq:frechetder1}
			D_V [{U}_\g]
			=
			{U}_\g \cdot \int\limits_0^1\! dt \ {U}_{\g^t}\cdot {\cal F}_{\m\n}(t) \[V^\m(t)\wedge \dot{\g}^\n(t) \] \cdot {U}_{\g^t}^{-1} \ , 
		\ee	
where 
\be
	{U}_{\g^t} = {\cal P} \exp{\left[i g \int\limits_{0}^t\! {\cal A}_\m(x)\ \dot{\gamma}^\m d\sigma \right]_\gamma} \ , \ x_\mu (\sigma) = \dot{\gamma}_\mu \sigma \  , \ \sigma \in [0,1] \ , \  x_\mu (0) = x_\mu (1)\ , \ {U}_{\g} = {U}_{\g^1} \ ,          
	\label{eq:paralleltransporter_0}
\ee
such that 
\be
\left(S_{12}\frac{ \delta}{\delta S_{12}} + S_{23}\frac{ \delta}{\delta S_{23}}\right) \  {\cal W}_\g  = D_{V_1} \ {\cal W}_\g  \ ,  \  		
{\cal W}_{\gamma}
=
\Big\langle 0 \Big| \frac{1}{N_c} {\rm Tr} \ {U}_\g \Big| 0 \Big\rangle \ .
\ee
Therefore, one gets the renormalization-group evolution in the form \cite{WL_UA} 
\begin{equation}
 \mu \frac{d}{d\mu} \ \left[ D_{V_1} \  {\cal W}_\g \right]
 =
- \sum \Gamma_{\rm cusp} \ ,
\label{eq:final}
\end{equation}
where $\Gamma_{\rm cusp}$ is the light-like cusp anomalous dimension \cite{KR87,WL_LC_all} and the summation runs over the number of cusps affected by the shape variation. 
The Fr\'echet derivatives for $D_{V_i}, i = 2,3,4$, which deliver other possible conformal transformations of the contour $\gamma$, can be constructed in the similar manner. 
This is not surprising since both operators induce the same shape variations that are shown in figure \ref{fig:WLQ_possible_area_var_LC}, but it  also justifies that the differential operators in (\ref{eq:newdiffop}) can be made mathematically well-defined. We also checked \cite{WL_UA} that the evolution equation (\ref{eq:final}) is valid (in the leading order) for a $\Pi$-shaped contour (figure \ref{fig:WLQ_pi_shape}) with finite light-like 
part \cite{LargeX_KM}. 	
This contour is a typical ingredient of the gauge-invariant operator expressions for various parton distribution functions and its geometric scaling is related to the rapidity evolution of the latter \cite{ChLargeX}.
	
At this point we would like to emphasize that one can go from the local area derivative, used by Makeenko and Migdal in their loop equations, to the Fr\'echet derivative \cite{ChMVdV_DeG}. This can easily be seen by identifying the two local vector fields, defining the infinitesimal shape variation of the area derivative, with the contour tangents and the (local) diffeomorphism generating vector field associated with the Fr\'echet derivative. If one now integrates over all these local variations along the entire contour, one arrives at the Fr\'echet derivative corresponding to the contour diffeomorphism. Moreover, it is worth mentioning that the infinitesimal version of  the Fr\'echet derivative (without the integration along the entire contour) has also been used by Polyakov (see Eqs. (3.5, 3.6) in Ref. \cite{Polyakov:1980ca}) without, however, establishing the relation to the Fr\'echet derivative. These issues will be addressed in a separate  work. 

In the present Letter we check, making use of the results existing in the literature and performing only trivial calculations, that our conjecture (\ref{eq:final}) is not only valid in the leading perturbative order, but can be extended beyond it (even up to the all orders for the quadrilateral in ${\cal N}=4$ super-Yang-Mills theory) due to the non-abelian exponentiation theorem \cite{WL_Exp},
and explicitly demonstrate this for the next-to-leading order $O(\alpha_s^2)$ using the two-loop result of Ref. \cite{Drummond:2007cf}.
Taking into account the running of the coupling constant, we check the validity of our conjecture in the next-to-leading order in the QCD case for the quadrilateral planar contour
and for the planar $\Pi$-shaped contour. In both cases we relied on the results presented in Refs. \cite{LargeX_KM,WL_LC_all}.

\section{Quadrilateral light-like Wilson loop in ${\cal N}=4$ SYM}\label{sec:two_loop_SYM_quadrilateral}

Let us consider the fully light-like contour $\gamma$ shown on Fig. \ref{fig:WLQ_possible_area_var_LC}. In the case of  the ${\cal N}=4$ SYM, the non-Abelian exponentiation theorem for the Wilson loops \cite{WL_Exp} allows us to present Wilson loop functionals in the form
				\be\label{eq:WQL_exponentiation}
					{\cal W}_\g 
					= 
					1+ \sum_{n=1}^\infty \lr{\frac{\alpha_S}{\pi}}^n {\cal W}^{(n)} 
					=
					\exp\left[{\sum_{n=1}^\infty \lr{\frac{\alpha_s}{\pi}}^n c^{(n)} w^{(n)}}\right]\  ,
				\ee
where ${\cal W}^{(n)}$ are the perturbative expansion terms of the Wilson loop, $c^{(n)}w^{(n)}$  the contribution to ${\cal W}^{(n)}$ corresponding to the ``maximally non-Abelian'' color factors $c^{(n)}$. From this we can write \cite{Drummond:2007cf}
			\be\label{exp-example}
				{\cal W}^{(1)} = 
				C_F w^{(1)} \ , \   {\cal W}^{(2)} = C_F N w^{(2)}+ \frac12C_F^2
				\lr{w^{(1)}}^2 \  \text{etc.} \ ,    
			\ee
which can be used to write the two-loop expression of the Wilson loop functional as:
				\be\label{W-decomposition}
					\ln {\cal W}_\gamma = \frac{\alpha_s}{\pi} C_F w^{(1)}  +  \lr{\frac{\alpha_s}{\pi}}^2 C_F C_A
					w^{(2)}  + {\cal O}({\alpha_s}^3)\  .
				\ee	
		The leading contributions read \cite{Drummond:2007cf} 
				\be\label{w1}
					w^{(1)}=-\frac{1}{\vep^2} \left[\lr{{- S_{12}}\,{\mu^2}}^\vep+\lr{
					{- S_{23}}\,{\mu^2}}^\vep\right]  + 
					\frac12 \ln^2\lr{\frac{S_{12}}{S_{23}}} + \frac{\pi^2}{3}+ {\cal O}(\epsilon)\ , 
				\ee
and
				\be\label{para1}
					w^{(2)} =  \left[(S_{12}\mu^2)^{2\epsilon}+
					(S_{23}\mu^2)^{2\epsilon}\right] \bigg\{ \epsilon^{-2} \frac{\pi^2}{48} +
					\epsilon^{-1}\frac{7}{8}\zeta_3\bigg\} -
					\frac{\pi^2}{24}\ln^2\lr{\frac{S_{12}}{S_{23}}} - \frac{37}{720}\pi^4 +
					{\cal O}(\epsilon)\ .  
				\ee
		Applying the differential operator (\ref{eq:newdiffop_1}) to this result returns:
			\be
				\shapeder_1 w^{(2)}  =
					\left[S_{12}\mu^2)^{2\epsilon}+
					(S_{23}\mu^2)^{2\epsilon}\right] \bigg\{ \epsilon^{-1} \frac{\pi^2}{24} +
					\frac{7}{4}\zeta_3\bigg\}+\text{finite} \ .  
			\ee
		Operating on this result with the mass scale derivative and taking the $\vep\to 0$ limit we finally arrive at:
			\be
				\m \frac{d}{d \m} {\shapeder}_1 \ w^{(2)}=
				4 \cdot \frac{\pi^2}{12}\  ,
			\ee	
		which becomes $4\cdot \frac{\pi^2}{12}C_F C_A $ when taking
		the non-Abelian color factors into account. Combining this with the one-loop result (\ref{w1}),
		again multiplying with the correct color factors, this returns:
			\be
			  \shapeder  
			  \ln {\cal W}_\gamma = 4 \cdot \lr{- \lr{\frac{\a_s}{\pi}} C_F+\lr{\frac{\a_s}{\pi}}^2C_F C_A  \frac{\pi^2}{12}} = - 4 
				\G_\text{cusp}\  ,
			\ee	
		consistent with our original conjecture (\ref{eq:final}) if one considers $\G_\text{cusp}$ as in \cite{Drummond:2007cf}, and where we took into account the $\frac{1}{N}$ factor in the definition of the Wilson loop
			\be
				\gcusp\lr{g} = \lr{\alpi}C_F - \lr{\alpi}^2C_F C_A  \frac{\pi^2}{12} + O (\alpha_s^4) \  .
			\ee	
		
%%%%%%%%%%%%%%%%%%%%%%%%%%%%%%%%%%%%%%%%%%%%%%%%%%%%%
\section{All order check in the ${\cal N}=4$ SYM theory}\label{sec:all_loop}

In \cite{Drummond:2007cf} it is also discussed that the Wilson loop can be split up in a divergent and a finite part:
				\be\label{cf.}
					\ln  {\cal W}_n  = \ln Z_n + \frac{1}{2} \Gamma_{\rm cusp}(a) F_n + {\cal O}(\epsilon)\  ,
				\ee
where $a = \frac{\alpha_s N_c}{Ž\pi}$ and the divergences are absorbed into the factor $Z_n$ and depends on the renormalization scale $\m$, $\vep$. In any gauge theory 
(see \cite{Sud_FF} and references therein) this factor can be written as\footnote{In Ref. \cite{Drummond:2007cf} $\gcusp$ is defined by means of the derivative $\frac{\pd}{\pd \ln{\m^2}}$, while we define it using $\frac{\pd}{\pd \ln{\m}}$, explaining
		the factor $2$ of difference in (\ref{eq:WLQ_allordereqZn})}:
				\begin{equation}\label{eq:WLQ_allordereqZn}
					\ln Z_n = - \half\sum_{l=1}^\infty a^l\left(\frac{\Gamma_{\rm
					cusp}^{(l)}}{(l\epsilon)^2} + \frac{\Gamma^{(l)}}{l\epsilon} \right)\sum_{i\neq j}^n
					({- x_{i,i+2}}\mu^{2})^{l\epsilon},
				\end{equation}
		where $\G_{\rm cusp}(a) = \sum_{l=1}^{\infty} a^l \G^{(l)}_{\rm cusp}$, $\G(a)
		= \sum_{l=1}^{\infty} a^l \G^{(l)}$ and $n$ is the number of cusps along the contour. The term $F_n$ refers to a finite contribution
		 that is parametrized only by the $x_i$ {(which are now combined in the $S_{ij}$)}, i.e. independent of the UV scale $\m$.
Then we get
				\begin{equation}\label{eq:WLQ_allordereqZn2}
					\ln Z_n = -\frac{1}{2}\sum_{l=1}^\infty a^l\left(\frac{\Gamma_{\rm
					cusp}^{(l)}}{(l\epsilon)^2} + \frac{\Gamma^{(l)}}{l\epsilon} \right)\sum_{i\neq j}^n
					(- S_{ij}\,\mu^{2})^{l\epsilon} \  .
				\end{equation}
		Applying the generalization, to $n$ segments, of the differential operator defined in (\ref{eq:newdiffop_1}) to (\ref{cf.}) returns
			\be
				\sum_{i} \shapeder_i \ln { {\cal W}_n} 
				=
					- \frac{1}{2}\sum_{l=1}^\infty a^l\left(\frac{\Gamma_{\rm
					cusp}^{(l)}}{(l\epsilon)} 
					+ 
					\Gamma^{(l)} \right)\sum_{i \neq j}^n
					(-S_{ij}\,\mu^{2})^{l\epsilon} + {\cal O}(\epsilon)\  .
			\ee
		Now again applying the ultraviolet scale derivative then returns
			\be
				\mu \frac{d}{d \mu}
				\sum_{i} \shapeder_i \  \ln {\cal W}_n =
				-\sum_{l=1}^\infty a^l\left(\Gamma_{\rm
					cusp}^{(l)} + l \epsilon \Gamma^{(l)} \right)\sum_{i\neq j}^n
					(- S_{ij}\,\mu^{2})^{l\epsilon} + {\cal O}(\epsilon)\  .
			\ee
		Taking the limit $\vep\to 0$ we get the final result:
			\be
				\m \frac{d}{d \mu} \  \sum_{i} \shapeder_i \ln {\cal W}_n 
				=
				- n \sum_{l=1}^\infty a^l\left(\Gamma_{\rm
					cusp}^{(l)} \right) 
					= 
					- n \G_{\rm cusp} 
					=  - \sum_\text{cusps}\gcusp\  ,
			\ee	
demonstrating that in  the {SYM} theory our conjecture holds to all orders. Of course, this result depends strongly on the non-Abelian exponentiation theorem and on the behaviour of the Sudakov form factor in this theory.
		
\section{Two loop QCD with the running coupling}\label{sec:two_loop_QCD}

In this Section we investigate the validity of our conjecture at the {NLO} level for the $\Pi$-shaped contour $\gamma_\pi$, Fig. \ref{fig:WLQ_pi_shape}, and quadrilateral on the contour Fig. \ref{fig:WLQ_possible_area_var_LC} from before, but now in QCD. In QCD we will need to take the running of the coupling constant into account, given that the $\beta$-function reads
		\be\label{eq:WQL_betafunction}
				\beta(g) = - \lr{\frac{11}{3}-\frac{2}{3}N_f} \frac{g^3}{16\pi^2} \ .
		\ee
	In other words, in a QCD setting beyond the leading order, the evolution equation is conjectured to be
		\be\label{eq:WQL_newconjecture}
				\lr{\mu \frac{\partial}{\partial \mu} + \beta (g) \frac{\partial}{\partial g}} \shapeder_1 \ln{\cal W}_\gamma 
				= 
				- \sum_\text{cusps} \gcusp \ .
		\ee	
The fact that we need to introduce this modification is not surprising, since $\gcusp$ depends on the coupling constant and hence is sensitive to its renormalization which is exactly described by the $\beta$-function. To demonstrate the validity we use the NLO result for the $\Pi$-shaped contour from Ref. \cite{LargeX_KM}.
Here the authors actually already proved (\ref{eq:WQL_newconjecture}) for this contour, with a bit of different notation\footnote{See eq. (4.4) in \cite{LargeX_KM} and the discussion below it.}, but the geometrical interpretation of the derivative $\frac{\partial }{\partial \lr{v\cdot y}}$ remained obscure. 
We shall explain this derivation in a bit more detail starting from the {NLO} expression for $\gamma_\pi$. This will show the strategy to follow to demonstrate the validity of (\ref{eq:WQL_newconjecture}) for the quadrilateral contour.
	
The renormalized {NLO}  expression for the $\Pi$-contour is given by  \cite{LargeX_KM}:
			\be\label{eq:WQL_NLO_PI}
				{\cal W}^{\rm R}_{\gamma_\pi} = \lr{\alpi} C_F\lr{-L^2 + L - \frac{5}{24}\pi^2} + \lr{\alpi}^2 C_F\lr{ A_1 L^3 + A_2 L^2 + A_3 L + {\cal O}(L^0)} 
				+ {\rm finite} \ ,  
			\ee
	where
			\ba
				A_1 & = &-\frac{11}{18}C_A+\frac{1}{9} N_f \  , \  
				A_2 = \left(\frac{1}{12}\pi^2-\frac{17}{18}\right)C_A
				       +\frac{1}{9} N_f  \  ,  \   
				\\
				A_3 & = &\left(\frac{9}{4}\zeta(3)-\frac{7}{18}\pi^2-\frac{55}{108}\right)C_A
				 +\left(\frac{1}{18}\pi^2-\frac{1}{54}\right) N_f \ ,  \label{res:re} \ ,   
						\ea
						where the new shape variable  $\tilde S_{12} = ({\tilde \ell_1} \cdot \ell_2)$ is introduced and 
						$$
						\tilde \ell_1^\mu = (\tilde \ell_1^+, \tilde \ell_1^-, \vecc 0_\perp) \ , \ \tilde \ell_1^2 \neq 0 \ ,  \  \ell_2^\mu = (0^+, \ell_2, \vecc 0_\perp)  \ , \  \ell_2^2 = 0,\ {L= \ln{\lr{S_{12} \m^2}}}  \ .    
						$$
Although the length of the non-light-like side is (semi-)infinite, this is irrelevant for the issues we are concerned {about} .  			
If we now consider the series coefficients of $L$, after the application of the differential operator  (\ref{eq:newdiffop_1}) and the mass scale derivative
to (\ref{eq:WQL_NLO_PI}), then it is easy to see that the coefficients of $L^2$ get multiplied to powers of $\lr{\alpi}$ higher than two, due to the presence of the $\beta$-function as multiplicative factor. Thus, they do not contribute at the NLO level. The coefficients of $L$, on the other hand, come from the $L^3$ term in Eq. (\ref{eq:WQL_NLO_PI})
by application of  
$$
\lr{\mu \frac{\partial}{\partial \mu}}\  \shapedertilde
$$
and from 
$$
\lr{ \beta (g) \frac{\partial}{\partial g}} \  \shapedertilde \lr{\alpi} C_F (-L^2) \ .  
$$
		Their total contribution becomes (where we use the notation: $\beta(g)=\beta  \frac{g^3}{16\pi^2}$):
			\be
				6A_1 - \beta 
				= 
				- \lr{\frac{11}{3}C_A-\frac{2}{3}N_f} +  \lr{\frac{11}{3}C_A-\frac{2}{3}N_f} = 0 \  .
			\ee
		If our conjecture is now to hold the constant terms should add up to $-2 \gcusp$ to order ${\cal O}\lr{\lr{\alpi}^3}$.
		The first contribution to the constant terms comes from $\lr{\alpi} C_F\lr{-L^2}$:
			\be
				\lr{\mu \frac{\partial}{\partial \mu}} \  \shapedertilde \lr{\alpi} C_F\lr{-L^2} = - 2 \lr{\alpi}C_F = -2\gcusp^\text{LO} \ .
			\ee
		The second contribution comes from the term $ \lr{\alpi}^2 C_F\lr{A_2 L^2}$:
			\ba
				\lr{\mu \frac{\partial}{\partial \mu}} \shapedertilde \lr{\alpi}^2 C_F\lr{A_2 L^2} &=&
				2 A_2 \lr{\alpi}^2 C_F\nn\\
				&=&2 \lr{\left(\frac{1}{12}\pi^2-\frac{17}{18}\right)C_A+\frac{1}{9} N_f} \lr{\alpi}^2 C_F \ .\nn\\
			\ea	
		Finally the third contribution comes from the term $\lr{\alpi} C_F\lr{ L}$:
			\be
				\lr{ \beta (g) \frac{\partial}{\partial g}} \  \shapedertilde \lr{\alpi} C_F (L)=
					\half \beta \  .
			\ee	
		The first term already contributes  to the cusp anomalous dimension in the correct way, so we only need to focus on the second and the third terms.
		Adding both contributions and extracting a $-2$ factor we get
		\ba
				-2 \lr{\alpi}^2 C_F \lr{-A_2 - \half \half \beta} &=&
				-2 \lr{\alpi}^2 C_F \lr{\left(-\frac{1}{12}\pi^2+\frac{17}{18}\right)C_A-\frac{1}{9} N_f + \frac{11}{12}C_A-\frac{2}{12}N_f}\nn\\
				&=&
				-2 \lr{\alpi}^2 C_F\lr{C_A\lr{\frac{67}{36}-\frac{1}{12}\pi^2} - \frac{5}{18}N_f}= -2 \gcusp^\text{NLO}\  .
			\ea
		Combining all the contributions shows that indeed (\ref{eq:WQL_newconjecture}) is valid at {NLO} for
		the $\gamma_\pi$ contour, with $\gcusp$ as in \cite{LargeX_KM}:
			\be
				\gcusp(g)=\alpi C_F+\left(\alpi\right)^2
				C_F\left(C_A\left(\frac{67}{36}-\frac{\pi^2}{12}\right)-N_f\frac5{18}\right)\  .
				\label{cusp}
			\ee

Let us now go back to the quadrilateral contour $\gamma$. To  proceed, we make use of  the {NLO} results for $\gamma$  derived in Refs. \cite{WL_LC_all}. 
In these papers it was shown that using the non-Abelian exponentiation theorem \cite{WL_Exp} {that } the renormalized quadrilateral Wilson loop can be written as
			\be
				{\cal W}_\gamma  
				= 
				\exp{\lr{{\cal W}_\gamma^{(1)} + {\cal W}_\gamma^{(2)}}}\  ,
			\ee
		where
			\ba
				{\cal W}_\gamma^{(1)} & = & - \frac{\a_s}{2\pi} C_F\lr{\ln^2{\lr{S_{12}\m^2}+\ln^2{\lr{S_{23}\m^2}}}}\label{eq:WQL_oneloopwlexp}\\	
				{\cal W}_\gamma^{(2)} & = & - \lr{\frac{\a_s}{\pi}}^2 C_F
					\left[
						w_1 \ln^3{\lr{S_{12}\m^2}}+ w_2 \ln^2{\lr{S_{12}\m^2}}+ w_3 \ln{\lr{S_{12}\m^2}}\ln{\lr{S_{23}\m^2}}  \right.  \\ 
						& + & \left. w_4 \ln{\lr{S_{12}\m^2}} 
						+  \lr{S_{12}\leftrightarrow S_{23}} + \text{const} \nn
						\right]\label{eq:WQL_twoloopwlexp}
			\ea	
		and with:
			\ba
				w_1 &=& \lr{\frac{11}{72}C_A-\frac{N_f}{36}} \ , \ 
				w_2 = \lr{\frac{67}{72}-\frac{\pi^2}{12}}C_A -\frac{5}{36}N_f\nn\\
				w_3 &=&  \lr{\frac{\pi^2}{24}C_A} \ , \  
				w_4 = \lr{\frac{101}{54} - \frac{7}{4}\zeta\lr{3} }C_A - \frac{7}{27}N_f \ .
			\ea	
		Note that (\ref{eq:WQL_oneloopwlexp}) is just another way to express $w^1$ from (\ref{w1}) in the renormalized version. To see how they are
		related, we take into account that
		$$ 
		z^\e =1 +\e \ln {z} + \half \e^2\ln^2{\lr{z}}+{\cal O}\lr{\e^3} 
		$$ 
		and recall that this is multiplied by a factor $\es$ in the LO result \cite{WL_UA}.  Subtracting the poles using the $\widebar{\text{MS}}$ scheme returns (\ref{eq:WQL_oneloopwlexp}) if one applies the same transformation to $ \lr{S_{23}\m^2}^\e$. We point out that applying our derivative followed by the mass scale derivative to (\ref{eq:WQL_oneloopwlexp}) gives again our conjecture at leading order:
			\be\label{eq:WQL_newconjLO1}
				\mu \frac{d}{d \mu}\  {\Bigg\langle\frac{\delta}{\delta \ln S}\Bigg\rangle_1  {{\cal W}_\gamma^{(1)}}} = - 8 \frac{\a_s}{2\pi} C_F = - 4\gcusp^\text{LO} \ .  
			\ee
		We want to do the same for the two loop result (\ref{eq:WQL_twoloopwlexp}), but just as for $Pi$-shaped contour we will
		need to change to our adapted conjecture (\ref{eq:WQL_newconjecture}).
		Doing this it is again easy to see that the $\ln^2$-terms after application of our generalized or Fr\'echet derivative, and the mass scale derivative,
		only contribute to higher orders of $\alpi$, i.e. to {NNLO} terms.
		Similarly to the $\Pi$-shaped contour case the terms contributing to the $\log$-terms cancel:
			\be
				 \half\beta +2 \cdot 6 w_1 = - \half\lr{\frac{11}{3}C_A - \frac{2}{3}N_F} + 12 \lr{\frac{11}{72}C_A - \frac{N_f}{36}} = 0 \ .  
			\ee
		The total contribution to the constant terms, after application of all the derivatives, is given by:
			\ba
				- 4\lr{\alpi}^2 C_F\lr{2w_2 +2 w_3} &=& - 4\lr{\alpi}^2 C_F\left[C_A\left(\frac{67}{36} - \frac{1}{12}\pi^2\right) - \frac{5}{18}N_f\right]\nn\\
				&=& \gcusp^\text{NLO}\  ,
			\ea
which combined with (\ref{eq:WQL_newconjLO1}) proves our conjecture at the {NLO} level for the quadrilateral light-like path $\gamma$. 

\section{Conclusion} 
We have shown that our original conjecture, Eq. (\ref{eq:final}), is valid in the ${\cal N}=4$ super-Yang-Mills theory, not only in the leading and next-to-leading orders of the perturbative expansion, but can be extended to all orders. The latter is possible since the $\beta$-function is zero in ${\cal N}=4$  SYM and there is no running of the coupling constant. On the other hand, working in QCD we demonstrated that taking into account the running of the strong coupling constant $\alpha_s$ allows us to verify this conjecture for
the $\Pi$-shaped and the quadrilateral light-like Wilson loops. Further development of the Fr\'echet derivative approach to the study of the geometrical and conformal properties of the polygonal Wilson loops with light-like segments will be reported elsewhere \cite{CM_2014}.

\end{document}